\documentclass[11pt]{article}
\usepackage{amsmath}
\usepackage{amsthm}
\usepackage{mathtools}
\usepackage{amssymb}
\usepackage{geometry}
\usepackage{cases}
\usepackage{color}
\usepackage{authblk}
\usepackage{listings}
\usepackage{float}
\usepackage{hyperref}
\usepackage{amsmath,amsfonts,amssymb,amsthm,mathtools}
\usepackage{latexsym,graphicx}
\usepackage{dsfont}
\usepackage{comment}
\usepackage{amscd}
\usepackage[dvipsnames]{xcolor}
\usepackage{extarrows} 
\usepackage{yfonts}
\usepackage{graphicx}
\usepackage{appendix}
\usepackage{caption}
\usepackage{subcaption}
\usepackage{verbatim}
\usepackage{authblk}
\usepackage{doi}

\newcommand{\ee}{\mathrm{e}}
\newcommand{\ii}{\mathrm{i}}

\newcommand{\T}{\mathbb{T}}
\newcommand{\C}{\mathbb{C}}

\newcommand{\D}{\mathbb{D}}
\newcommand{\Z}{\mathbb{Z}}

\usepackage{xcolor}
\definecolor{riverlane_green}{RGB}{0, 111, 98}
\definecolor{riverlane_light_green}{RGB}{0, 150, 143}
\definecolor{riverlane_orange}{RGB}{255, 117, 0}
\definecolor{riverlane_red}{RGB}{220, 68, 5}
\definecolor{riverlane_pink}{RGB}{207, 111, 127}
\usepackage{hyperref}
\hypersetup{
  colorlinks   = true, 
  urlcolor     = riverlane_green, 
  linkcolor    = riverlane_orange, 
  citecolor   = riverlane_green  
}

\def\Xint#1{\mathchoice
{\XXint\displaystyle\textstyle{#1}}%
{\XXint\textstyle\scriptstyle{#1}}%
{\XXint\scriptstyle\scriptscriptstyle{#1}}%
{\XXint\scriptscriptstyle\scriptscriptstyle{#1}}%
\!\int}
\def\XXint#1#2#3{{\setbox0=\hbox{$#1{#2#3}{\int}$}
\vcenter{\hbox{$#2#3$}}\kern-.5\wd0}}

\def\pvint{\Xint-}

\newtheorem{problem}{Problem}
\newtheorem{theorem}{Theorem}
\newtheorem{corollary}{Corollary}

\newtheoremstyle{break}
  {\topsep}{\topsep}%
  {\itshape}{}%
  {\bfseries}{}%
  {\newline}{}%
\theoremstyle{break}

\newcommand\blfootnote[1]{%
  \begingroup
  \renewcommand\thefootnote{}\footnote{#1}%
  \addtocounter{footnote}{-1}%
  \endgroup
}

\title{Two exact quantum signal processing results}

\begin{document}

\author[1,2]{Bjorn K. Berntson}
\author[2]{Christoph S\"{u}nderhauf}
\affil[1]{Riverlane Research,
Cambridge, Massachusetts}
\affil[2]{Riverlane,
Cambridge, United Kingdom}

\date{May 19, 2025}

\maketitle

\begin{abstract}
Quantum signal processing (QSP) is a framework for implementing certain polynomial functions via quantum circuits. To construct a QSP circuit, one needs (i) a target polynomial $P(z)$, which must satisfy $\lvert P(z)\rvert\leq 1$ on the complex unit circle $\mathbb{T}$ and (ii) a complementary polynomial $Q(z)$, which satisfies $\lvert P(z)\rvert^2+\lvert Q(z)\rvert^2=1$ on $\mathbb{T}$. We present two exact mathematical results within this context. First, we obtain an exact expression for a certain uniform polynomial approximant of $1/x$, which is used to perform matrix inversion via quantum circuits. Second, given a generic target polynomial $P(z)$, we construct the complementary polynomial $Q(z)$ exactly via integral representations, valid throughout the entire complex plane. 
\end{abstract}

\section{Introduction}
\blfootnote{Emails: bjorn.berntson@riverlane.com, christoph.sunderhauf@riverlane.com}
Quantum signal processing (QSP) is a novel representation of polynomials by a product of unitary matrices, which may be implemented directly on quantum computers \cite{low2017}. Crucially, QSP may be generalized to the quantum singular value transformation (QSVT), which allows for these same polynomial functions to be applied to matrices \cite{gilyen2019}. Usually, we are interested in implementing polynomials which are uniform approximants of certain non-polynomial functions. For example, by approximating the function $1/x$ on an appropriate subset of $[-1,1]$ excluding $0$, we may perform approximate matrix inversion using the QSVT. Approaches to implementing polynomial approximants within the QSP/QSVT framework often rely on various approximations, such as Chebyshev truncations or computation of complementary polynomials via numerical methods. While standard approximation theoretic and numerical techniques are often effective, exact results provide a deeper insight into the structure of the problem and solution and, in the present setting, reduce the computational resources, quantum or classical, required to solve problems. 

\section{Matrix inversion polynomials}

The first problem we address is related to the numerical inversion of matrices via the QSVT. Suppose we are given a matrix such that the singular values are contained in $[a,1]$ for some $a\in (0,1)$. Let $S(a)\coloneqq [-1,-a]\cup [a,1]$.  A key step in inverting the given matrix via the QSVT is to solve the following problem. 

\begin{problem}[Constructing the matrix inversion polynomial]\label{prob:1}
For $a\in (0,1)$ and $n\in \Z_{>0}$, find $P_{2n-1}\in \mathbb{R}[x]$, $\deg(P_{2n-1})=2n-1$ such that
\begin{equation}
\varepsilon_{2n-1}(a)\coloneqq \bigg\lVert P_{2n-1}(x)-\frac{1}{x}\bigg\rVert_{\infty,S(a)}
\end{equation} 
is minimized. 
\end{problem}

Problem~\ref{prob:1} has been solved by Privalov \cite{privalov2007}. Our contribution is an explicit form for the optimal polynomial, together with an explicit form for the associated error $\varepsilon_{2n-1}(a)$. 

Recall that the Chebyshev polynomials of the first kind are defined by $T_n(\cos(\theta))=\cos(n\theta)$ for  $n\in \mathbb{Z}_{\geq 0}$. Let
\begin{equation}
L_n(x;a)\coloneqq \frac{1}{2^{n-1}}\bigg( T_n(x)+\frac{1-a}{1+a}T_{n-1}(x)\bigg). 
\end{equation}
Our result is the following. 
\begin{theorem}
The minimizing polynomial in Problem~\ref{prob:1} can be expressed as 
\begin{equation}
P_{2n-1}(x;a)=\begin{dcases}\frac{1}{x}-\frac{L_n\big(\frac{2x^2-(1+a^2)}{1-a^2};a\big)}{xL_n\big(-\frac{1+a^2}{1-a^2};a\big)} & x\in \mathbb{R}\setminus\{0\} \\
0 & x=0.
\end{dcases}
\end{equation}
The associated error is given by
\begin{equation}
\varepsilon_{2n-1}(a)=\frac{\sqrt{1+a^2}}{a}\frac{(1-a)^n}{(1+a)^{n-1}}.
\end{equation}
\end{theorem}

This result allows for significantly expedited classical pre-processing of matrix inversion polynomials versus the standard Remez method. Additionally, the optimality of $P_{2n-1}$ guarantees a minimal circuit depth versus approximants obtained with the Remez method.

\section{Complementary polynomials}

Generalized quantum signal processing \cite{motlagh2024}, which contains standard QSP parameterizations as special cases, has demonstrated the fundamental importance of complementary polynomials. More specifically, it is mathematically trivial to implement a target polynomial\footnote{In this section, different from the last, we consider polynomials on the complex unit circle. The correspondence with polynomials on $[-1,1]$ is obtained using the identity $T_n(x)=\frac12(z^n+z^{-n})$ for $x=\mathrm{Re}\,z$ and the scaling symmetry $P(z)\to z^n P(z)$ for any $n\in \Z$.} $P(z)$ once the following problem is solved.

\begin{problem}[Complementary polynomials problem]\label{prob:CP}
Given $P\in \C[z]$ satisfying $\lvert P(z)\rvert\leq  1$ for $z\in \mathbb{T}\coloneqq \{z\in \mathbb{C}:\lvert z\rvert =1\}$, find $Q\in \mathbb{C}[z]$ such that $\deg(Q)=\deg(P)$ and
\begin{equation}\label{eq:PQ}
\lvert P(z)\rvert^2+\lvert Q(z)\rvert^2=1 \quad (z\in \mathbb{T}) 	
\end{equation}
holds.
\end{problem}

Several numerical approaches to obtaining complementary polynomials have been proposed in the literature \cite{motlagh2024, gilyen2019, ying2022}. Here, we construct an exact representation of solutions to Problem~\ref{prob:CP} as contour integrals and develop a numerical algorithm to obtain $Q$ explicitly in the monomial basis. Additionally, we use our exact representation of $Q$ to perform error analysis and obtain the scaling in $\varepsilon$, the desired accuracy in the supremum norm.

Let
\begin{equation}\label{eq:P}
P(z)=\sum_{n=0}^d p_nz^n \quad (p_0\neq 0);
\end{equation}
the restriction that $p_0\neq 0$ is imposed without loss of generality as $\lvert z^n P(z)\rvert=\lvert P(z)\rvert$ holds for all $z\in \T$, $n\in \Z$. Our main theoretical result is the following theorem. 

\begin{theorem}\label{thm:main}
Suppose $P(z)$ satisfying the assumptions of Problem~\ref{prob:CP} is given in the form \eqref{eq:P}. Let $d_0\in \Z_{\geq 0}$ be the number of roots of $1-\lvert P(z)\rvert^2$ on $\T$, not counting multiplicity, and $\{(t_j,2\ell_j)\}_{j=1}^{d_0}$ be the corresponding roots and multiplicities, which are necessarily even. Then, for $z\in \D\coloneqq \{z\in \C:\lvert z\rvert< 1\}$,
\begin{subequations}\label{eq:Q}
\begin{equation}
Q(z)=
{Q}_0(z)\exp\Bigg(\frac{1}{4\pi\ii}\int_{\T} \frac{z'+z}{z'-z}R(z)	\frac{\mathrm{d}z'}{z'}\Bigg) ,\label{eq:QD}
\end{equation}
for $z\in \T$, 
\begin{equation}
Q(z)={Q}_0(z)\exp\Bigg(\frac{1}{4\pi\ii}\,\pvint_{\T}\frac{z'+z}{z'-z}R(z) \frac{\mathrm{d}z'}{z'}+\frac12R(z) \Bigg) ,\label{eq:QT} 
\end{equation}
and for $z\in \C\setminus\overline{\D}$, 
\begin{equation}
Q(z)=\frac{1-P(z)P^*(1/z)}{Q_0^*(1/z)}\exp\Bigg(\frac{1}{4\pi\ii}\int_{\T}\frac{z'+z}{z'-z}R(z)\frac{\mathrm{d}z'}{z'}\Bigg), \label{eq:QCD}
\end{equation}
\end{subequations}
where
\begin{equation}
R(z)\coloneqq \log \bigg(\frac{1-\lvert P(z')\rvert^2}{\lvert {Q}_0(z')\rvert^2}\bigg)
\end{equation}
and
\begin{equation}\label{eq:Qtilde}
{Q}_0(z)\coloneqq \prod_{j=1}^{d_0} (z-t_j)^{\ell_j}
\end{equation}
and the integration contour $\T$ is positively-oriented, and the dashed integral indicates a principal value prescription with respect to the singularity $z'=z$ on $\T$, solves \eqref{eq:PQ}. Moreover, \eqref{eq:Q} is, up to a multiplicative phase, the unique solution of \eqref{eq:PQ} with no roots in $\D$. 

\end{theorem}

Theorem~\ref{thm:main} provides an exact representation of $Q$ in Problem~\ref{prob:CP}. To obtain $Q$ explicitly in the monomial basis, it suffices to evaluate \eqref{eq:Q} at any $d+1$ points of $\C$ and employ the Lagrange interpolation formula. Our numerical approach is based on interpolation through roots of unity; this is equivalent to a discrete Fourier transform. The following corollary of Theorem~\ref{thm:main} establishes a Fourier analytic variant of the integral representation \eqref{eq:QT} of $Q$ on $\T$, which may be used to evaluate $Q$ at the roots of unity.\footnote{We remark that an equation similar to \eqref{eq:QFourier} is contained in \cite{magland2005}.}

\begin{corollary}\label{cor:main}
The representation \eqref{eq:QT} of $Q$ on $\T$ is equivalent to ($\theta\in (-\pi,\pi]$)
\begin{equation}\label{eq:QFourier}
Q(\ee^{\ii\theta})={Q}_0(\ee^{\ii\theta})\exp\Bigg( \Pi \bigg[\log\bigg(\frac{1-\lvert P(\ee^{\ii\theta})\rvert^2}{\lvert{Q}_0(\ee^{\ii\theta})\rvert^2}\bigg)\bigg] \Bigg),
\end{equation}
where the Fourier multiplier $\Pi$ is defined by
\begin{equation}\label{eq:Piexp}
\Pi[\ee^{\ii n \theta}]=\begin{cases}
\ee^{\ii n\theta} & n\in \Z_{> 0} \\
\frac12 & n=0 \\
0 & n\in \Z_{<0}.	
\end{cases}	
\end{equation}
\end{corollary}

Following from Corollary~\ref{cor:main} is an algorithm to construct $Q$ by means of the Fast Fourier Transform. While a detailed description of this algorithm is outside the scope of the present article, we can report that for a polynomial $P$ with known $\lVert P(z)\rVert_{\infty,\T}<1$, this algorithm scales as $N=O(\log(1/\varepsilon))$, where $N$ is the dimension of the discrete Fourier basis and 
\begin{equation}
\varepsilon\coloneqq \big\lVert Q(z)-\tilde{Q}_N(z) \big\rVert_{\infty,\T},
\end{equation}
with $\tilde{Q}_N(z)$ the approximation of $Q(z)$ obtained from our algorithm. We emphasize that, versus existing numerical methods based on either root-finding \cite{gilyen2019}, Prony's method \cite{ying2022}, or objective function minimization \cite{motlagh2024}, our algorithm is amenable to explicit error analysis. 

\section*{Acknowledgment}

This work was partially funded by Innovate UK (Grant reference 10071684). We thank Earl Campbell, Tanuj Khattar, Nicholas Rubin, Tharon Holdsworth, and especially Hari Krovi for useful discussions. We additionally thank Marius Bothe, Zalan Nemeth, and Andrew Patterson for collaboration on closely related projects.

\bibliographystyle{unsrt}
\bibliography{two_exact_QSP.bib}

\end{document}